# Capitalizing Risk, Regulation, and Revised Perceptions: Sequential Shocks in Florida's Condominium Market

Shaoming Cheng
Department of Public Policy and Administration
Florida International University
scheng@fiu.edu

**[Abstract]** Housing markets capitalize new information about future risks and ownership costs, yet less is known about how capitalization evolves when closely sequenced information and regulatory shocks alter what markets had previously priced. This study examines Florida's condominium market following the Surfside collapse and the subsequent enactment of Senate Bill 4-D (SB4D). Surfside heightened the salience of structural and financial risks, while SB4D established a new regulatory regime involving inspections, reserve requirements, and related ownership obligations for qualifying condominium buildings. Using statewide administrative condominium transaction data from 2020 to 2024, the analysis employs a common-treatment, three-period difference-in-differences design comparing condominiums in buildings inferred to meet the statutory three-story threshold with those in lower-rise buildings across the pre-Surfside, post-Surfside transition, and post-SB4D periods. The results show statistically significant, negative capitalization at the three-story threshold during the post-Surfside transition, with an estimated relative price decline of approximately 3.4% in the adjusted specification. Negative capitalization is significantly greater in the post-SB4D regime, reaching approximately 6.1%. Heterogeneity results further show greater post-SB4D negative capitalization among younger and higher-valued properties, with weaker evidence for properties closest to the coast. These patterns are consistent with capitalization reflecting revisions in perceived future user costs relative to what had already been incorporated into property values, rather than the absolute level of those costs alone. A policy-anticipation analysis further divides the post-Surfside transition into an immediate post-collapse period and a pre-enactment period in which regulatory intervention was increasingly anticipated. Capitalization does not differ significantly between these two periods, suggesting that the greater negative capitalization in the post-SB4D period is less likely to reflect a progressively intensifying post-Surfside effect. The findings show how capitalization following sequential information and regulatory shocks depends on the extent to which prior perceptions are revised.

## Capitalizing Risk, Regulation, and Revised Perceptions: Sequential Shocks in Florida's Condominium Market

### Introduction

Property values adjust as housing markets capitalize new information about future risks and ownership costs. Much of the housing capitalization literature examines repeated shocks of the same type, such as successive hurricanes, floods, or earthquakes, in which later events reinforce or update perceptions of a common underlying hazard. Less is known about how housing markets respond when closely sequenced shocks differ fundamentally in nature. An unexpected information shock may revise perceptions of structural or financial risk, while a subsequent regulatory intervention may alter expected ownership costs by introducing new legal obligations. Although both mechanisms may affect housing prices, they represent conceptually distinct channels of capitalization that are rarely observed separately. The magnitude of capitalization should depend on how much each shock changes perceived future risks and ownership costs beyond what was already reflected in prices.

Florida's condominium market provides a distinctive setting in which to examine these sequential information and regulatory shocks. The collapse of Champlain Towers South in Surfside on 24 June 2021 heightened the salience of structural integrity, deferred maintenance, reserve adequacy, insurance exposure, and condominium governance. Less than one year later, Florida enacted Senate Bill 4-D (SB4D), establishing milestone-inspection and Structural Integrity Reserve Study requirements and related reserve obligations for qualifying condominium buildings. Surfside therefore represented an unexpected information shock capable of changing perceptions of condominium risk and anticipated user costs, whereas SB4D subsequently introduced a regulatory regime that made future inspection, reserve, repair, and related ownership obligations more certain. The close temporal succession of these events provides an unusual setting for examining how capitalization changes as an initial information shock is followed by regulatory intervention.

Existing research documents important condominium-market responses to Surfside. Beracha et al. (2026) find that older condominiums in Miami-Dade County experienced larger post-collapse price discounts and weaker market liquidity than younger condominiums, with stronger effects among taller buildings. Building height is particularly relevant to the present study because SB4D's statutory inspection and reserve requirements apply to condominium buildings of three stories or more. The analysis therefore uses the three-story threshold to examine capitalization within the condominium market. A three-period difference-in-differences (DID) model compares condominiums in buildings of three stories or more with those in lower-rise buildings across the post-Surfside transition and subsequent SB4D regulation.

The Surfside–SB4D setting also relates to a growing DID literature on multiple treatments and evolving policy regimes. Research on multiple treatments cautions that coefficients from successive treatments may not represent isolated causal effects (Fricke, 2017; de Chaisemartin & D'Haultfœuille, 2023; Goldsmith-Pinkham, Hull, & Kolesár, 2024). Related work examines policies that evolve through connected stages. Ann et al. (2026), for example, study a housing policy that progressed from announcement through implementation and subsequent modification, allowing effects to vary across policy stages. Although Surfside is an information shock rather than a policy treatment, a similar interpretive issue arises here. The post-SB4D period follows both the Surfside shock and the post-Surfside transition period, which includes a potential policy-anticipation phase when information about possible regulatory action may have begun to affect expectations before SB4D was formally enacted. Accordingly, the period-specific coefficients measure capitalization at the three-story threshold relative to the common pre-Surfside period rather than isolated causal effects of Surfside and SB4D.

The analysis uses statewide administrative property assessment and condominium transaction records maintained by the Florida Department of Revenue (DOR), covering all 67 Florida counties from 2020 through 2024. Because SB4D applies at the building level and establishes a three-story statutory threshold, the treatment group consists of condominium units in buildings inferred to have three or more stories, with units in lower-rise buildings forming the comparison group. A monthly, three-period DID design compares the pre-Surfside period, the post-Surfside transition period, and the post-SB4D period while maintaining a common treatment definition throughout. This design allows capitalization at the three-story threshold to vary across the sequential information and regulatory regimes without redefining the treatment group. Monthly event-study estimates trace the evolution of the treatment effect and assess pre-treatment dynamics. A separate policy-anticipation specification examines whether the greater negative capitalization observed during the post-SB4D period was already developing during the post-Surfside transition before SB4D was formally enacted.

The empirical results reveal a clear difference in capitalization across the two post-event regimes. Following Surfside, condominiums in buildings meeting the three-story threshold experienced significant negative capitalization relative to lower-rise condominiums, with an estimated relative price decline of approximately 3.4% in the adjusted specification. Negative capitalization was significantly greater in the post-SB4D regime, reaching approximately 6.1%. This pattern indicates that repricing at the three-story threshold emerged during the post-Surfside transition and became more pronounced under the subsequent regulatory regime. Heterogeneity results further show greater post-SB4D negative capitalization among higher-valued properties and the newest buildings, with weaker evidence for properties closest to the coast. These patterns are consistent with capitalization reflecting not simply the level of perceived risk or expected ownership costs, but the extent to which prior perceptions are revised. Older buildings, for example, may face higher maintenance, reserve, repair, insurance, and structural costs in absolute terms, but some of these disadvantages may already have been reflected in property values. The additional ownership costs associated with the post-SB4D regime may therefore represent a smaller revision in prior perceptions for older buildings than for the newest buildings, even if older buildings continue to face higher expected costs in absolute terms. A policy-anticipation analysis further divides the post-Surfside transition into an immediate post-collapse period and a pre-enactment period in which regulatory intervention was increasingly anticipated. The analysis finds no significant difference in capitalization between these two periods, suggesting that the greater negative capitalization in the post-SB4D regime is less likely to reflect a progressively intensifying post-Surfside effect.

This study makes four contributions. First, it extends the housing capitalization literature by examining two closely sequenced but fundamentally different shocks: an unexpected information shock that heightened the salience of structural and financial risks and a subsequent regulatory regime that altered anticipated ownership obligations. The setting highlights a central feature of capitalization under sequential shocks: price adjustments depend not simply on the level of perceived risk or expected ownership costs, but on the extent to which prior perceptions are revised. This distinction helps explain why properties facing higher risks or costs in absolute terms need not experience the largest price adjustments following a new information or regulatory shock. Second, the study identifies capitalization at SB4D's statutory three-story threshold within the condominium market. By comparing condominiums in buildings meeting the three-story threshold with those in lower-rise buildings, the analysis isolates a within-condominium margin directly related to the law's inspection and reserve requirements while leaving broader repricing common to both groups outside the treatment effect. Third, the study incorporates insights from the literature on multiple treatments and evolving policy regimes by estimating capitalization at a common treatment threshold across successive post-event periods. This approach recognizes that the post-SB4D regime follows both the Surfside information shock and a post-Surfside transition during which regulatory intervention became increasingly anticipated, and therefore does not interpret differences between regime-specific coefficients as isolated causal effects of each event. Finally, the statewide design demonstrates how administrative property assessment and transaction records can be used to infer building-level regulatory exposure and conduct large-scale, transparent, and reproducible analysis of housing-market responses to regulatory change.

## Institutional Background of SB4D

On 24 June 2021, the partial collapse of the Champlain Towers South condominium in Surfside, Florida, claimed 98 lives and prompted widespread concern regarding the safety, maintenance, and governance of aging condominium buildings. At the federal level, the National Institute of Standards and Technology (NIST) launched a comprehensive technical investigation to determine the causes of the collapse. In June 2026, following a five-year investigation, NIST concluded that the collapse originated with the failure of two garage column–pool deck connections, which initiated progressive structural failures over several weeks before the catastrophic collapse (NIST, 2026). Although the investigation identified the physical mechanism responsible for the failure, the tragedy heightened public attention to structural deterioration, deferred maintenance, reserve adequacy, and condominium governance.

The collapse also prompted extensive policy reviews by professional organizations, policymakers, and industry stakeholders. The *Report of the Florida Bar RPPTL Condominium Law and Policy Life Safety Advisory Task Force* (2021) identified systemic concerns related to deferred maintenance, aging building infrastructure, inadequate reserve funding, and weaknesses in condominium governance. The Task Force recommended strengthening structural inspection requirements and improving long-term financial planning for major structural components, concluding that structural safety and capital replacement were too important to depend solely on owner discretion. As Schwartz and Koushel (2022) observed, condominium associations have historically been permitted to waive or reduce reserve funding, and owners frequently elected to postpone maintenance, repairs, and capital replacement because of the immediate financial burden.

Responding to these concerns, the Florida Legislature enacted SB4D during a special legislative session in May 2022. SB4D introduced sweeping reforms affecting condominium governance, structural oversight, and reserve funding. The legislation established two principal requirements. First, it created mandatory milestone inspections for condominium buildings three stories or higher, requiring an initial inspection at 30 years of age, or 25 years for buildings located within three miles of the coastline, followed by recurring inspections every ten years thereafter. These inspections must be performed by licensed engineers or architects to identify substantial structural deterioration and other conditions affecting structural safety. Second, SB4D required condominium associations to prepare SIRS and to fund reserves in accordance with those studies. For specified structural components, condominium associations could no longer waive or reduce reserve contributions, fundamentally changing long-standing financial practices governing condominium ownership.

Implementation of SB4D presented practical and administrative challenges. Condominium associations, engineers, local governments, and professional organizations raised concerns regarding inspection capacity, compliance deadlines, reserve funding requirements, and the financial burden on condominium owners. In response, the Florida Legislature enacted Senate Bill 154 (SB154) in 2023, commonly referred to as the "Glitch Bill," to clarify and modify several provisions of the original legislation. Among other changes, SB154 revised milestone-inspection deadlines, modified reserve requirements, and provided greater flexibility in implementing the new regulatory framework. The Legislature made additional changes through House Bill 1021 (HB1021), effective 1 July 2024, which further revised condominium governance, milestone-inspection, and reporting requirements. These amendments collectively reflect the continued implementation and refinement of the regulatory framework established by SB4D during the study period.

**Capitalization Following Information and Regulatory Shocks**

Capitalization theory posits that housing prices reflect the discounted present value of expected future housing services net of expected ownership costs (Oates, 1969; Poterba, 1984; Rosen, 1974). Property values therefore incorporate anticipated future amenities, structural quality, maintenance obligations, insurance premiums, regulatory requirements, and other recurring ownership expenses. When new information or regulation changes perceptions of future risks or ownership costs, those revisions are capitalized into property prices. Importantly, the resulting price adjustment depends not simply on the level of perceived risk or expected user costs, but on the extent to which prior perceptions are revised. Properties with relatively high expected user costs may already trade at lower prices because those disadvantages have been anticipated and capitalized, leaving less additional adjustment when new information arrives.

Conversely, properties previously perceived as facing lower risks or costs may experience larger price adjustments if a shock produces a greater revision in those perceptions. The extent of capitalization may also depend on local market conditions, with stronger capitalization generally observed where housing supply is constrained by regulatory or geographic factors (Hilber, 2016).

The housing literature on natural hazards provides substantial evidence for this mechanism. Flood and hurricane risks are capitalized into residential property values, with major disasters often increasing discounts as previously underappreciated risks become salient (Bin & Polasky, 2004; Hallstrom & Smith, 2005). These effects may subsequently diminish as salience fades or perceptions change (Bin & Landry, 2013; Bui et al., 2024). Davlasheridze et al. (2026) similarly find that the effects of Hurricane Harvey varied across property-value levels and largely dissipated within approximately one year. Variation in the magnitude and persistence of capitalization has also been documented following earthquakes (Murdoch et al., 1993; Naoi et al., 2009), tornadoes (Cohen & Gutkowski, 2026), long-run exposure to sea-level rise (Soni, 2026), and other climate-related hazards (Hino & Burke, 2021). Behavioral evidence provides a complementary explanation for these responses: low-probability hazards may be underweighted until a salient event changes perceptions of the likelihood or consequences of future losses, while myopia, fading memories, and limited awareness can prevent objective risk from being fully reflected in prices (Sutter & Poitras, 2010; Fletcher et al., 2022). Consistent with this interpretation, Hurricane Sandy affected property values even among homes without direct physical damage (Cohen et al., 2021), distant hurricanes increased flood-risk discounts in markets without direct storm impacts (Fang et al., 2023), and Hurricane Harvey increased capitalization of probability-based flood-hazard information beyond conventional floodplain designations (Akinlotan et al., 2026).

Greater underlying exposure, however, need not produce a larger price response when that exposure was already recognized and capitalized. Gibson and Mullins (2020), for example, find that the post-Sandy price effect for flooded properties already located within the new floodplain was not statistically distinguishable from zero, while Ellen and Meltzer (2024) document persistent effects concentrated in particular portions of the housing market, including properties outside pre-existing flood zones. Together with evidence that disaster effects vary across property-value levels (Davlasheridze et al., 2026), these findings reinforce a central implication of capitalization theory: price adjustments following new information reflect not simply the underlying level of exposure, but the extent to which prior perceptions of future risks and ownership costs are revised.

This distinction becomes especially relevant when information and regulatory shocks occur sequentially because the first shock changes the pricing baseline against which the subsequent regime is evaluated. Beracha et al. (2026), for example, find larger post-Surfside price and liquidity effects for older relative to younger condominiums. The present setting instead examines capitalization at the three-story threshold as the market moves from the Surfside information shock to the SB4D regulatory regime. Once some structural and financial risks have been incorporated into condominium prices following Surfside, the response to SB4D should depend on the extent to which the new regime further revises market perceptions of future risks and ownership costs. This logic also provides a basis for heterogeneous responses across property types. Older buildings, for example, may face greater maintenance, reserve, repair, insurance, and structural costs in absolute terms, but these disadvantages may already be reflected, at least partly, in property values. The additional user costs associated with SB4D may therefore represent a smaller revision for older buildings than for younger buildings previously perceived as facing fewer major building-related expenditures.

The same capitalization mechanism operates when new information or mitigation reduces anticipated future losses. Hazard-mitigation investments can increase property values when buyers perceive that they credibly reduce expected damages and ownership costs, particularly when those investments are observable or independently verified. Professional inspections, engineering evaluations, and certification programs can reduce information asymmetry by revealing otherwise difficult-to-observe structural characteristics and allowing resilience investments to enter property valuations (Gatzlaff et al., 2018). Homes certified under the FORTIFIED Home Program command measurable price premiums consistent with capitalization of lower expected storm losses and insurance costs (Awondo et al., 2023; Petrolia et al., 2023), while flood-protection infrastructure and other protective measures can similarly generate price premiums following major storms and mitigation investments (Davlasheridze & Fan, 2019; Kalhoro et al., 2026). This evidence underscores that capitalization can reflect expected benefits as well as expected costs. In the present setting, capitalization at the three-story threshold therefore reflects the net valuation of anticipated regulatory obligations and any expected benefits from greater structural oversight and resilience.

Housing regulation provides a related channel through which future costs, risks, and benefits are incorporated into property values. Regulations can affect housing supply, construction quality, information availability, and future operating or compliance costs (Gyourko & Molloy, 2015). At the property level, two mechanisms are particularly relevant. First, regulation can reduce information asymmetry by making otherwise difficult-to-observe characteristics more credible. Residential energy-performance disclosure requirements, for example, increase the capitalization of energy efficiency by revealing information relevant to future operating costs (Frondel et al., 2020; Myers et al., 2022), while engineering inspections can facilitate capitalization of otherwise hidden structural mitigation features (Gatzlaff et al., 2018). Second, regulation can directly alter anticipated costs and losses through construction standards, inspection requirements, reserve obligations, and related legal requirements. Building and energy codes consequently affect housing quality and property values (Muzio et al., 2025). Homes constructed under stricter wildfire building codes in high-hazard areas command premiums consistent with greater structural resilience (Wrenn, 2026), while hurricane-resistant building standards similarly generate price premiums when buyers capitalize expected reductions in future losses (Dumm et al., 2011, 2012). SB4D combines these channels by introducing information-producing inspection requirements alongside reserve and related obligations that can alter future ownership costs.

Other recurring ownership expenses reinforce the same capitalization mechanism. Higher homeowners' insurance premiums reduce housing prices because they increase expected ownership costs and may also convey information about underlying risk (Koning, 2019; Nyce et al., 2015). Homeowners' association fees are similarly capitalized as recurring components of housing user costs (Narwold et al., 2018), while unresolved building-code violations reduce transaction prices by signaling building deficiencies and anticipated compliance expenditures (Bartram, 2019). Together, these studies show that property values respond not simply to current expenditures or observable physical conditions, but to revisions in the anticipated stream of costs and benefits associated with ownership.

A related methodological literature emphasizes that settings involving multiple treatments or evolving policy stages require greater interpretive care than a conventional single-treatment

DID. Fricke (2017) develops identification results for DID settings with multiple treatments, showing that comparisons among treatment states generally require assumptions beyond those of a standard treated-versus-untreated design. More recent work demonstrates that regressions containing multiple treatment indicators can combine heterogeneous effects across treatments, complicating causal interpretation of individual coefficients (de Chaisemartin & D'Haultfœuille, 2023; Goldsmith-Pinkham, Hull, & Kolesár, 2024). Goldsmith-Pinkham et al. (2024), in particular, show that coefficients in multiple-treatment regressions can contain non-convex weighted contributions from the effects of other treatments. Related work on sequential housing policies illustrates how a common treatment margin can operate across distinct policy stages. Ann et al. (2026), for example, examine a housing policy that progresses from announcement through implementation and subsequent modification, explicitly modeling effects across successive regimes. This literature motivates allowing capitalization at a fixed treatment margin to vary across regimes while avoiding interpretation of differences between regime-specific coefficients as isolated causal effects.

The Surfside–SB4D setting brings these theoretical and methodological considerations together within a common condominium-market comparison. Surfside was not a policy treatment but an unexpected information shock that could revise perceptions of structural, governance, and financial risk across the condominium market. SB4D subsequently established a regulatory regime that made inspection, reserve funding, repair, and related obligations more certain for buildings meeting the statutory three-story threshold. The empirical question is whether capitalization at that threshold changed as the condominium market moved from the post-Surfside transition to the post-SB4D regime. Because the post-SB4D period follows an information shock that may already have revised market perceptions, and regulatory intervention may have been anticipated before formal enactment, the two post-event coefficients are not interpreted as cleanly separable causal effects. The relevant economic object is instead capitalization at the three-story threshold within each regime relative to the common pre-Surfside baseline. This interpretation motivates the three-period design and the separate policy-anticipation identification check.

## Research Questions and Hypotheses

The Surfside condominium collapse and subsequent enactment of SB4D provide an opportunity to examine two sequential but distinct shocks within a common capitalization framework. Both events could revise prior perceptions, but through different channels: Surfside primarily heightened perceptions of structural, governance, and financial risk, while SB4D established a regulatory regime that made inspection, reserve, repair, and related ownership obligations more certain for qualifying buildings. The analysis examines how capitalization at the three-story threshold changed as perceptions were revised following these sequential shocks. This study addresses three research questions.

RQ1. Did condominiums in buildings meeting the three-story threshold experience capitalization relative to lower-rise condominiums following the Surfside collapse?

RQ2. Did capitalization relative to lower-rise condominiums change during the post-SB4D regulatory regime?

RQ3. Did these capitalization patterns vary across condominium value, building age, and coastal proximity?

Capitalization theory predicts that housing prices adjust when new information or regulation revises prior perceptions of future risks or ownership costs. The magnitude of that adjustment depends not simply on the level of perceived risk or expected costs, but on the extent of the revision relative to what was already capitalized. Properties facing greater risks or costs may therefore experience smaller price adjustments if those disadvantages were already reflected in property values, while previously less-exposed properties may experience larger adjustments when perceptions change more substantially. Accordingly, capitalization following Surfside and SB4D may vary at the three-story statutory threshold and across condominium characteristics according to the extent to which each shock revises prior perceptions.

H1. Holding other factors constant, following the Surfside condominium collapse, condominiums in buildings meeting the three-story threshold are expected to experience negative capitalization relative to lower-rise condominiums as prior perceptions of structural, governance, and financial risks are revised.

H2. Holding other factors constant, during the post-SB4D period, condominiums in buildings meeting the three-story threshold are expected to experience negative capitalization relative to lower-rise condominiums as prior perceptions of future ownership costs are revised in response to mandatory inspections, reserve funding requirements, and related regulatory obligations.

The extent of these revisions may differ across condominium characteristics and therefore produce heterogeneous capitalization at the three-story threshold.

H3a (Condominium Value). Holding other factors constant, capitalization at the three-story threshold is expected to vary across condominium value segments as Surfside and SB4D revise prior perceptions of future risks and ownership costs.

H3b (Building Age). Holding other factors constant, capitalization at the three-story threshold is expected to vary across building-age segments as Surfside and SB4D revise prior perceptions of future risks and ownership costs. These differences are expected to reflect the extent of the revision rather than the level of building-related risks or costs alone.

H3c (Distance to Coast). Holding other factors constant, capitalization at the three-story threshold is not expected to vary systematically with coastal proximity because Surfside and SB4D primarily concern building-related risks and regulatory obligations rather than coastal exposure.

## Empirical Strategy

The empirical analysis employs a three-period difference-in-differences (DID) design to estimate within-condominium capitalization at the three-story threshold across two sequential, closely related events affecting Florida's condominium market, namely, the Surfside condominium collapse and the subsequent enactment of SB4D. The design compares condominiums in buildings with three or more stories with those in lower-rise buildings across three periods: the pre-Surfside period, the post-Surfside transition period, and the post-SB4D period. Both post-event periods are evaluated relative to the same pre-Surfside baseline using a common treatment definition based on SB4D's three-story threshold. Maintaining the same comparison groups and baseline across periods allows capitalization at that threshold to be compared across the two sequential events without changing the underlying treated population. The analysis further evaluates whether this treatment effect varies across condominium value, building age, and coastal proximity. A separate policy-anticipation specification examines whether the greater negative capitalization in the post-SB4D period reflects a post-Surfside decline that was already intensifying before enactment. As a sensitivity check, the main specification is also re-estimated using the 12 months immediately preceding Surfside as the pre-Surfside period to assess robustness to the choice of pre-event window.

***Three-Period Difference-in-Differences***

The pre-Surfside period extends from January 2020 through May 2021, the post-Surfside transition period from June 2021 through May 2022, and the post-SB4D period from June 2022 through December 2024. The close temporal proximity of Surfside and SB4D creates an identification challenge because the regulatory response developed in the aftermath of the collapse. A simple pre/post-SB4D specification could conflate changes during the intervening post-Surfside transition with those observed under the subsequent regulatory regime, while a simple pre/post-Surfside specification would not distinguish between the two post-event periods. The three-period design addresses this issue by estimating capitalization at the three-story threshold separately for the post-Surfside transition and post-SB4D periods, with both estimates evaluated relative to the same pre-Surfside baseline. The baseline specification is:

$$\boldsymbol{\ln(P_{ibct}) = \beta_1(\mathrm{PostSurfside}_t \times \mathrm{Treat}_i) + \beta_2(\mathrm{PostSB4D}_t \times \mathrm{Treat}_i) + X_{it}\gamma + \alpha_b + \delta_c + \tau_t + \varepsilon_{ibct}}$$

Here, $ln(P_{ibct})$ is the natural logarithm of the sales price for condominium unit $i$ in building $b$, Census Block $c$, and year-month $t$. $Treat_i$ identifies condominium units in buildings classified as three stories or higher under the treatment construction procedure described in the previous section. $PostSurfside_t$ equals one for transactions occurring after the Surfside collapse but before the enactment of SB4D, while $PostSB4D_t$ equals one for transactions occurring from June 2022 through December 2024 occurring following the enactment of SB4D. The omitted category is the pre-Surfside period from January 2020 through May 2021. The coefficient $\beta_1$ estimates capitalization of three-story-and-higher condominiums relative to lower-rise condominiums during the post-Surfside transition period, while $\beta_2$ estimates the corresponding capitalization during the SB4D period, in each case relative to the pre-Surfside baseline.

The model includes building fixed effects ($\alpha_b$), census-block fixed effects ($\delta_c$), and year-month fixed effects ($\tau_t$). Building fixed effects absorb time-invariant differences across condominium buildings, including location, amenities, construction quality, height, and other persistent structural characteristics. Census Block fixed effects account for highly localized neighborhood conditions, while year–month fixed effects absorb common temporal shocks, including statewide housing-market trends, seasonality, interest-rate movements, insurance-market conditions, and other macroeconomic influences. The vector $X_{it}$ includes transaction-level controls in the adjusted specifications.

Standard errors are two-way clustered by building and year-month to account for within-building correlation in transaction prices and contemporaneous shocks affecting transactions within the same month. Because the relevant event dates are common across units rather than staggered across treatment cohorts, identification does not rely on variation in treatment timing. A monthly event-study specification complements these estimates by tracing relative capitalization at the three-story threshold over time and assessing pre-event dynamics.

***Heterogeneity Analysis***

To examine whether capitalization at the three-story threshold varies across building characteristics, the three-period DID is extended along three dimensions, i.e., condominium value, building age, and coastal proximity. All models retain the covariates, fixed effects, and clustering structure of the adjusted main specification.

For condominium value, groups are defined using the statewide distribution of condominium Just Value[1] in the 2021 property stock. The thresholds are fixed before Surfside so that subsequent transaction activity does not determine group membership. The bottom 50% serves as the common reference group and is compared with the top 50%, top 25%, and top 10% of the distribution. For building age, buildings 30 years or older serve as the common reference group and are compared with buildings under 10, under 20, and under 30 years old. The under-10 and under-20 specifications exclude buildings falling between the upper bound of the comparison group and 30 years, whereas the under-30 specification partitions the full age distribution. For coastal proximity, condominiums located 1.25 miles or more from the coast serve as the common reference group and are compared with condominiums located within 0.1, 0.5, and 1.25 miles of the coast. The 0.1- and 0.5-mile specifications exclude properties between the upper bound of the comparison group and 1.25 miles, whereas the 1.25-mile specification partitions the full distance distribution.

[1] Just Value is the Florida property appraiser's estimate of a property's market value for ad valorem tax purposes. It is distinct from assessed value and taxable value, which may reflect assessment limitations and exemptions. For more details, https://floridarevenue.com/Property/Pages/Taxpayers.aspx.

## Data and Variables

### *Data Sources*

The empirical analysis uses administrative property assessment and transaction records maintained by the Florida Department of Revenue (DOR). Specifically, the study combines the annual Name-Address-Legal (NAL) assessment roll with the Sales Data File (SDF), which together provide parcel-level property characteristics and recorded real estate transactions for all 67 Florida counties. These statewide administrative data have been used in prior research on Florida housing markets and property taxation (Allen & Dare, 2009; Ihlanfeldt & Mayock, 2014)

Compared with proprietary Multiple Listing Service (MLS) databases, the DOR files offer several advantages for this statewide analysis. They provide standardized coverage across all Florida counties, stable parcel identifiers that facilitate longitudinal matching, annual property assessments, and recorded transaction information. The availability of parcel-level Just Value is especially useful for defining pre-Surfside condominium value groups and for controlling for property value in the adjusted specifications. Together, these features make the DOR data well suited for examining within-condominium capitalization across Florida over time.

The analysis uses condominium transactions from 2020 through 2024, spanning the pre-Surfside period, the post-Surfside transition period, and the SB4D regulatory period. Property characteristics, including Just Value, effective year built, and parcel identifiers, are obtained from the NAL assessment files, while sales prices and transaction dates are obtained from the SDF. Coastal proximity is calculated by geocoding each parcel and measuring the straight-line distance to the nearest point on Florida's official shoreline using GIS shoreline data provided by the Florida Fish and Wildlife Conservation Commission.

***Sample Construction***

The analysis is restricted to residential condominium transactions identified using the Florida DOR property use classification and classified as arm's-length sales using the Sale Qualification Codes reported in SDF [2]. Transactions with missing or nonpositive sales prices are excluded. To limit the influence of extreme observations while preserving local market variation, sales prices are trimmed within each county and calendar quarter at the 1st and 99th percentiles. This localized procedure accommodates substantial geographic and temporal variation in Florida condominium prices without imposing a common statewide cutoff.

Two trimming procedures were evaluated for condominium transactions from 2020 through 2024: one based on sales price and the other on price per square foot, with both applying county-quarter-specific 1st and 99th percentile thresholds. The primary analyses use sales-price trimming because it provides the most transparent and directly interpretable criterion. This procedure retains 780,656 of 793,874 transactions (98.3%), excluding 13,218 observations (1.7%). Among transactions for which price per square foot can be constructed, the alternative procedure retains 766,402 observations (98.2%). The similar retention rates indicate that sample retention is not materially sensitive to the choice of trimming metric.

[2] Residential condominiums are identified using DOR Use Code (DOR_UC) 4. Arm's-length transactions are identified in the SDF using Sale Qualification Codes (QUAL_CD) 1 through 6.

***Variable Construction and Transformations***

The dependent variable is the natural logarithm of condominium sales price. Because the sample is restricted to transactions with positive sales prices, no offset is required before transformation. Building age is measured as the difference between the transaction year and the property's effective year built, with negative values set to zero. Coastal proximity is measured as the shortest straight-line distance between each condominium parcel and the nearest mapped Florida shoreline using GIS shoreline data from the Florida Fish and Wildlife Conservation Commission. Just Value and distance to the coast are transformed using $\ln(1 + X)$, because both may legitimately equal zero. The adjusted specifications include fixed 2021 Just Value, building age at sale, and distance to the coast as covariates. Using fixed 2021 Just Value provides a predetermined measure of property value that precedes both Surfside and SB4D and avoids conditioning on contemporaneous assessed values that may themselves reflect post-event capitalization.

***Treatment Variable Construction***

SB4D applies at the condominium-building level rather than the individual-unit level, so the treatment indicator is constructed at the building level. Under Florida law, milestone inspection and SIRS requirements apply to condominium buildings that are three stories or higher, while lower-rise buildings generally fall outside the statutory threshold. Accordingly, condominium units located in buildings with three or more stories are classified as treated, while units in buildings with fewer than three stories form the comparison group. This definition creates the within-condominium treatment margin used throughout the empirical analysis. The three-story threshold is maintained across the pre-Surfside, post-Surfside transition, and post-SB4D periods. Using a common treatment definition preserves the same treated and comparison groups and allows capitalization at the three-story threshold to be compared consistently across periods.

Building height is not consistently reported in the statewide administrative property tax records. Treatment status is therefore constructed from condominium unit information in the Florida Department of Revenue NAL file, rather than from the SDF transaction sample. The NAL covers the universe of condominium units, so building height can be inferred from all available units within a building rather than only those that happened to sell during the study period. This avoids incorrectly classifying a multi-story building as having fewer than three stories simply because the observed transactions occurred in lower-floor units. Unit designations are identified from both the property address and the Short Legal Description (LEGAL_N) field, which captures unit information even when it does not appear in the address.

A transparent, rule-based parsing procedure identifies common unit-numbering conventions, including simple numeric units (e.g., 603), alphanumeric identifiers (e.g., 401A, 10-E), letter-prefixed units (e.g., C-205, N-1021), and penthouse designations (e.g., PH, LPH, UPH, and DPH). Parsable unit designations are translated into inferred floor locations based on their numbering structure. For example, unit 603 is inferred to be on the sixth floor, 401A on the fourth floor, 10-E on the tenth floor, and C-205 on the second floor. Penthouse designations provide direct evidence that a building exceeds the three-story threshold. Unit-level information is then aggregated to the building level using a canonicalized building address that removes trailing unit identifiers while preserving the underlying building address and standardizing common address abbreviations. The procedure is designed to preserve distinct buildings that may have similar unit-numbering patterns. For example, units at 41114 FISHER ISLAND DR are grouped separately from those at 41116 FISHER ISLAND DR. Within each building, the highest inferred floor observed across parsable unit designations is used to determine whether the building crosses the three-story threshold. If the available unit records establish the presence of a third or higher floor, all condominium units associated with that building are classified as treated.

Condominium units without available unit information are classified as having fewer than three stories and retained in the comparison group. The absence of unit information (e.g., 1213 N Duval St and 6351 Pelican Bay Blvd) is more consistent with a single-story or two-story condominium configuration, where vertically differentiated unit numbering is less common. This approach involves a tradeoff between potential misclassification and sample exclusion. Any resulting classification error is more likely to place units in a three-story-or-higher building in the comparison group than to classify units in a lower-rise building as treated, because treatment requires observed unit information indicating a third or higher floor. To the extent that misclassified higher-rise units experience similar negative capitalization, their inclusion in the comparison group would attenuate the estimated treatment effect toward zero. The estimated capitalization effect would therefore be conservative with respect to this source of misclassification. Excluding all units without available unit information, however, would substantially reduce sample coverage and could alter the composition of the comparison group by disproportionately excluding lower-rise condominiums. Retaining these observations therefore preserves the broader condominium comparison while acknowledging the potential for some attenuation from treatment misclassification.

## Empirical Results and Findings

### *Baseline Capitalization Effects*

Table 1 reports the three-period DID estimates of condominium price capitalization during the post-Surfside transition and post-SB4D periods, with January 2020 through May 2021 serving as the omitted pre-Surfside period. The baseline specification includes building, Census Block, and year-month fixed effects, while the adjusted specification additionally controls for fixed 2021 Just Value, building age at sale, and distance to the coast. During the post-Surfside period, the estimated treatment effect is −0.0335 (SE = 0.0092) in the baseline specification and −0.0347 (SE = 0.0091) in the adjusted specification, with both estimates statistically significant at the 0.1 percent level. The adjusted estimate indicates that, during the post-Surfside period, transaction prices for condominiums in buildings with three or more stories were approximately 3.4 percent lower than those for lower-rise condominiums, relative to the pre-Surfside period.

[Insert Table 1 about here]

The negative capitalization becomes larger during the post-SB4D period. The estimated treatment effect is −0.0601 (SE = 0.0075) in the baseline specification and −0.0629 (SE = 0.0078) in the adjusted specification. The adjusted coefficient corresponds to approximately 6.1 percent lower transaction prices for condominiums in buildings with three or more stories relative to those in lower-rise buildings. The similarity of the baseline and adjusted estimates in both post-event periods indicates that the estimated capitalization is not explained by the included differences in property value, building age, or coastal proximity.

The post-Surfside and post-SB4D treatment effects are also statistically distinguishable. Wald tests reject equality of the two coefficients in the baseline ($p = 0.022$) and adjusted ($p = 0.018$) specifications. Thus, negative capitalization is evident following Surfside and becomes significantly greater during the post-SB4D period. This difference is not interpreted as a pure causal effect of SB4D, but it establishes a distinct empirical pattern across the two post-event periods. Figure 1 complements these period-average estimates by presenting monthly treatment effects relative to May 2021, the month immediately preceding the Surfside collapse. The event-study specification traces the evolution of capitalization at the three-story threshold over time and provides a visual assessment of the pre-Surfside dynamics relevant to the DID design.

[Insert Figure 1 about here]

The pre-Surfside monthly estimates generally fluctuate around zero without a clear directional pattern. Consistent with the visual evidence, the linear pre-trend tests do not reject the null of no systematic trend in either the baseline ($p = 0.098$) or adjusted ($p = 0.054$) specification, although the latter approaches conventional significance. Following Surfside, the monthly estimates fluctuate around a negative period average and become more negative on average during the post-SB4D period. Together, the event-study and three-period DID results show negative capitalization during the post-Surfside period and significantly greater negative capitalization during the post-SB4D period.

***Heterogeneous Capitalization***

Table 2 examines whether capitalization at the three-story threshold varies across condominium value, building age, and coastal proximity. Each column reports a separate DID specification in which the post-period treatment effects are interacted with an indicator for the relevant comparison group. The three-way interaction coefficients test whether, holding the other modeled factors constant, capitalization at the three-story threshold differs between the specified comparison and reference groups. All specifications retain the covariates, fixed effects, and clustering structure of the adjusted main model.

[Insert Table 2 about here]

Condominium Value. The value results indicate that heterogeneity during the SB4D period is concentrated toward the upper end of the condominium value distribution. Value groups are defined using the statewide distribution of condominium Just Value in the 2021 property stock, with the bottom 50 percent serving as the common reference group. The top 10 percent, top 25 percent, and top 50 percent are examined separately against this reference group. During the post-Surfside transition period, the three-way interactions are −0.0288 for the top 10 percent, −0.0190 for the top 25 percent, and 0.0170 for the top 50 percent, and none is statistically significant. The estimates therefore provide no clear evidence that capitalization at the three-story threshold differed systematically across condominium value segments during the transition period.

A different pattern emerges during the SB4D period. The three-way interaction is negative and statistically significant for the top 10 percent (−0.0511, SE = 0.0170, $p < 0.01$) and the top 25 percent (−0.0341, SE = 0.0158, $p < 0.05$), but is small and statistically insignificant for the top 50 percent (0.0116, SE = 0.0169). Relative to the bottom half of the value distribution, these estimates indicate more negative capitalization at the three-story threshold among condominiums in the upper quartile and especially the upper decile during the SB4D period. The insignificant interaction for the top half indicates that this pattern is concentrated in the upper tail rather than reflecting a general high-value effect.

The magnitude of the estimates illustrates this concentration. In the top-decile specification, the estimated SB4D-period treatment effect for the bottom-50-percent reference group is −0.0437, while the additional −0.0511 interaction implies an estimated treatment effect of approximately −0.095 log points for the top decile. In the top-quartile specification, the corresponding estimates are −0.0484 for the bottom-half reference group and approximately −0.083 log points for the top quartile after incorporating the interaction. These estimates are consistent with larger revisions in anticipated future ownership costs among higher-value three-story-and-higher condominiums relative to lower-rise condominiums, although the analysis does not identify the specific cost components underlying the capitalization response. The value thresholds also show that the result is not confined to an ultra-luxury segment: in the statewide 2021 condominium stock, the top quartile begins at approximately $252,000 in Just Value and the top decile at approximately $438,000. The stronger negative capitalization therefore extends across a meaningful portion of the upper end of Florida's condominium value distribution.

Building Age. Building-age heterogeneity is more concentrated. Buildings 30 years or older serve as the common reference group and are compared separately with buildings under 10, under 20, and under 30 years old. The broader age comparisons provide little evidence of systematic heterogeneity. When buildings under 30 years old are compared with those 30 years or older, the post-Surfside three-way interaction is −0.0021 and the SB4D-period interaction is 0.0012; neither is statistically significant. Restricting the comparison group to buildings under 20 years produces the same general result, with interactions of −0.0051 during the post-Surfside period and −0.0111 during the SB4D period, neither statistically distinguishable from zero. The estimates therefore do not support a general younger-versus-older gradient in capitalization.

The pattern changes when the comparison is restricted to the newest buildings. For condominiums in buildings under 10 years old, the post-Surfside three-way interaction is −0.0331 (SE = 0.0196), which is marginally significant at the 10 percent level. During the SB4D period, the interaction is −0.0600 (SE = 0.0201, $p < 0.01$). In this specification, the estimated SB4D-period treatment effect for the 30-years-or-older reference group is −0.0620, implying an estimated treatment effect of approximately −0.122 log points for buildings under 10 years old after incorporating the interaction. Age-related heterogeneity is therefore concentrated among the newest three-story-and-higher condominiums rather than increasing systematically across progressively younger age classifications.

This pattern is consistent with capitalization reflecting the extent to which prior perceptions of future ownership costs are revised rather than their absolute levels. Older buildings may already have incorporated expectations of greater maintenance, repair, reserve, and structural expenditures before Surfside and SB4D. The absence of significant differences in the broader under-20 and under-30 comparisons is therefore important: the results do not indicate that younger buildings generally experienced more negative capitalization. Instead, the strongest difference is concentrated among buildings less than 10 years old, for which the SB4D regulatory regime may have represented a larger revision relative to previously more favorable perceptions of future ownership costs.

Coastal Proximity. Coastal proximity provides little evidence of systematic heterogeneity in capitalization at the three-story threshold. Condominiums located 1.25 miles or more from the coast serve as the common reference group and are compared separately with properties located within 0.1 mile, within 0.5 mile, and within 1.25 miles of the coast. None of the post-Surfside three-way interactions is statistically significant. The estimated interactions are −0.0075 for properties within 0.1 mile, 0.0018 for those within 0.5 mile, and 0.0087 for those within 1.25 miles. These estimates therefore provide no evidence that capitalization at the three-story threshold varied systematically with coastal proximity during the post-Surfside transition period.

During the SB4D period, the three-way interactions are −0.0290 for properties within 0.1 mile of the coast, −0.0201 for those within 0.5 mile, and −0.0115 for those within 1.25 miles.

The interaction for properties within 0.1 mile is marginally significant at the 10 percent level (SE = 0.0171), while the broader coastal comparisons are statistically insignificant. Although the point estimates become more negative as the comparison is restricted to properties nearest the coast, the evidence does not indicate a systematic coastal gradient in capitalization at the three-story threshold. Taken together, the heterogeneity results show that significant SB4D-period differences are concentrated among higher-value condominiums and, separately, among buildings less than 10 years old, while coastal proximity provides little evidence of systematic heterogeneity. The average negative capitalization during the SB4D period therefore masks meaningful differences in particular condominium segments rather than a uniform pattern across all three dimensions.

***Sensitivity and Identification Checks***

Two complementary analyses assess the identification and robustness of the main findings. First, a policy-anticipation check examines whether the greater negative capitalization in the post-SB4D period reflects a post-Surfside decline that was already intensifying before enactment. Second, the analysis is re-estimated using the 12 months immediately preceding Surfside to assess robustness to the choice of pre-Surfside period.

Policy Anticipation as an Identification Check. A potential identification concern is that the greater negative capitalization observed during the post-SB4D period reflects a progressively intensifying post-Surfside effect rather than a change associated with the subsequent SB4D regulation. To assess this alternative explanation, Table 3 divides the post-Surfside transition into an immediate post-Surfside period (June–September 2021) and a policy-anticipation period (October 2021–May 2022), beginning with the October 2021 release of the *Report of the Florida Bar RPPTL Condominium Law and Policy Life Safety Advisory Task Force*. Because anticipated regulation may be capitalized before implementation (Ann et al., 2026), this division provides a substantively motivated test of whether negative capitalization was already intensifying before SB4D was enacted. The four-period specification retains the treatment definition, pre-Surfside reference period, and model structure of the main analysis.

[Insert Table 3 about here]

The results do not indicate a progressively intensifying post-Surfside effect. In the adjusted specification, the treatment effect is −0.0300 during the immediate post-Surfside period and −0.0374 during the policy-anticipation period, and the difference is not statistically significant ($p = 0.737$). The post-SB4D estimate remains −0.0628, essentially unchanged from the three-period specification, and the difference from the immediate post-Surfside estimate is statistically significant ($p = 0.024$). These findings provide little evidence that the greater negative capitalization in the post-SB4D period reflects the continuation of an intensifying post-Surfside effect.

12-Month Pre-Surfside Period Sensitivity Check. To assess the robustness of the main findings to the choice of pre-Surfside period, the analysis is re-estimated using the 12 months immediately preceding the collapse (June 2020–May 2021). The main analysis begins in January 2020, a natural starting point given the annual structure of the underlying administrative data, but there is no ex ante basis for specifying when a pre-event trend, if present, should begin. The alternative window is also equal in duration to the 12-month post-Surfside transition period (June 2021–May 2022).

[Insert Figure 2 about here]

[Insert Table 4 about here]

Figure 2 shows that the pre-Surfside estimates are generally concentrated around zero and do not display an evident progressively negative trajectory approaching Surfside. As shown in Table 4, the linear pre-trend is statistically insignificant in the baseline specification ($p = 0.166$) and in the adjusted specification ($p = 0.140$). These results provide stronger evidence of stable pre-treatment differences over the alternative pre-Surfside window. The three-period estimates remain highly stable. In the adjusted specification, the post-Surfside and post-SB4D estimates are −0.0313 and −0.0608, compared with −0.0347 and −0.0629 in the main specification. The difference between the two post-event estimates also remains statistically significant ($p = 0.012$). Overall, the main findings are robust to the alternative 12-month pre-Surfside period. The estimated capitalization is consistent with the main specification, and the greater negative capitalization in the post-SB4D period remains statistically significant relative to the post-Surfside period.

## Discussion and Conclusion

This study examines capitalization in Florida's condominium market following two sequential events: the Surfside condominium collapse and the subsequent enactment of SB4D. Using a common pre-Surfside baseline and a three-period DID framework, the analysis compares condominiums in buildings with three or more stories with those in buildings with fewer than three stories during the post-Surfside transition and post-SB4D periods. Negative capitalization is observed at the three-story threshold in both periods, but it is significantly greater in the post-SB4D period. In the adjusted specification, transaction prices are approximately 3.4 percent lower during the post-Surfside period and 6.1 percent lower during the post-SB4D period for condominiums in buildings with three or more stories relative to those in lower-rise buildings. The difference between the two estimates is statistically significant.

The post-Surfside estimate captures relative capitalization within the condominium market rather than the overall effect of the collapse on condominium prices. Any repricing common to condominiums would affect both three-story-and-higher and lower-rise properties and therefore would not be captured by the treatment contrast. The findings thus show that capitalization following Surfside also occurred within the condominium market, as revised perceptions of risk were reflected differently across building heights. Notably, this capitalization occurred at the building-height margin that subsequently became central to Florida's regulatory response under SB4D.

The heterogeneity results further illustrate the potential role of prior perceptions in capitalization. Housing prices reflect revisions in expected future risks and ownership costs relative to what market participants had already anticipated, rather than simply their absolute levels. During the post-SB4D period, negative capitalization at the three-story threshold is significantly greater among condominiums in the top 25 percent and top 10 percent of the statewide 2021 Just Value distribution. Building-age heterogeneity is less systematic. Broader comparisons involving buildings under 20 and under 30 years old show little evidence of significant differences, while buildings less than 10 years old exhibit substantially greater negative capitalization relative to buildings 30 years or older. One possible interpretation is that maintenance, repair, reserve, and structural expenditures were already more heavily reflected in the prices of older buildings, whereas the subsequent regulatory requirements represented a larger revision to prior perceptions for the newest buildings. Coastal proximity, by contrast, does not significantly modify the treatment effect. These patterns are consistent with capitalization reflecting revisions in perceived future risks and ownership costs rather than simply the underlying level of structural or financial exposure.

The sensitivity and identification checks support the robustness of the main findings to alternative temporal definitions. Dividing the post-Surfside transition into an immediate post-Surfside period and a policy-anticipation period beginning in October 2021 produces adjusted treatment estimates of −0.0300 and −0.0374, respectively, which are not statistically distinguishable. The post-SB4D estimate remains −0.0628, essentially unchanged from the main specification, and is statistically significantly more negative than the immediate post-Surfside estimate. Thus, the results do not indicate a progressively intensifying post-Surfside effect that would explain the greater negative capitalization observed in the post-SB4D period. The findings are also robust to using the 12 months immediately preceding Surfside as the alternative pre-Surfside period. The adjusted post-Surfside and post-SB4D estimates remain similar to the main estimates at −0.0313 and −0.0608, respectively, and the post-SB4D estimate remains statistically significantly more negative than the post-Surfside estimate.

The findings have broader implications for housing regulation. Policies intended to improve building safety and resilience can also affect property values as regulatory obligations revise perceptions of future ownership costs. In the case of SB4D, inspection and reserve requirements may generate important safety and informational benefits, while the associated obligations may also be capitalized into condominium prices. These price adjustments should not themselves be interpreted as the social costs of regulation, since they may reflect both anticipated costs and expected benefits. The heterogeneity across property value and building age further suggests that these effects are not uniform across the condominium market, raising distributional and equity considerations regarding how the financial consequences of safety regulation are borne across properties and owners. These considerations should be evaluated alongside the intended safety and resilience objectives of housing regulation.

**Table 1: Monthly Three-Period DID Estimates**

| DV: natural logged sales price | | |
|---|---|---|
| | **Baseline** | **Adjusted** |
| Post-Surfside × Treated | −0.0335*** | −0.0347*** |
| | (0.0092) | (0.0091) |
| Post-SB4D × Treated | −0.0601*** | −0.0629*** |
| | (0.0075) | (0.0078) |
| 2021 Fixed Just Value (log) | | 0.2873*** |
| | | (0.0414) |
| Building Age at Sale | | 0.0045 |
| | | (0.0073) |
| Distance to Coast (log) | | 4.560 |
| | | (2.766) |
| | | |
| Building Fixed Effects | Yes | Yes |
| Census Block Fixed Effects | Yes | Yes |
| Year–Month Fixed Effects | Yes | Yes |
| Observations | 734,204 | 704,651 |
| Overall $R^2$ | 0.9147 | 0.9207 |
| Within $R^2$ | 0.0010 | 0.0695 |
| | | |
| Linear pre-trend test (p-value) | 0.098 | 0.054 |
| Wald test (p-value): $H_0$: Post-Surfside = Post-SB4D | 0.022 | 0.018 |

Notes:

- The omitted (reference) period is Pre-Surfside (January 2020–May 2021). Post-Surfside denotes June 2021–May 2022, and Post-SB4D denotes June 2022–December 2024.
- Standard errors are reported in parentheses and are two-way clustered at the building and year–month levels.
- *** $p < 0.001$; ** $p < 0.01$; * $p < 0.05$.

## Table 2: Heterogeneity in Condominium Price Capitalization

| | **Just Value** | | | **Building Age** | | | **Coastal Proximity** | | |
|---|---|---|---|---|---|---|---|---|---|
| **Reference** | Bottom 50% | Bottom 50% | Bottom 50% | 30+ Yrs | 30+ Yrs | 30+ Yrs | ≥ 1.25 Miles | ≥ 1.25 Miles | ≥ 1.25 Miles |
| **Comparison** | Top 10% | Top 25% | Top 50% | < 10 Yrs | < 20 Yrs | < 30 Yrs | < 0.1 Miles | < 0.5 Miles | < 1.25 Miles |
| | | | | | | | | | |
| **Post-Surfside × Treated** | −0.0055 | −0.0162 | −0.0452* | −0.0328** | −0.0284* | −0.0338** | −0.0421† | −0.0415† | −0.0417† |
| | (0.0172) | (0.0170) | (0.0173) | (0.0121) | (0.0109) | (0.0105) | (0.0241) | (0.0239) | (0.0242) |
| **Post-SB4D × Treated** | −0.0437** | −0.0484*** | −0.0700*** | −0.0620*** | −0.0567*** | −0.0633*** | −0.0459** | −0.0471** | −0.0539** |
| | (0.0130) | (0.0128) | (0.0132) | (0.0098) | (0.0093) | (0.0081) | (0.0165) | (0.0162) | (0.0162) |
| | | | | | | | | | |
| **Post-Surfside × Treated × Comparison** | −0.0288 | −0.0190 | 0.017 | −0.0331† | −0.0051 | −0.0021 | −0.0075 | 0.0018 | 0.0087 |
| | (0.0194) | (0.0202) | (0.0211) | (0.0196) | (0.0129) | (0.0113) | (0.0261) | (0.0255) | (0.0256) |
| **Post-SB4D × Treated × Comparison** | −0.0511** | −0.0341* | 0.0116 | −0.0600** | −0.0111 | 0.0012 | −0.0290† | −0.0201 | −0.0115 |
| | (0.0170) | (0.0158) | (0.0169) | (0.0201) | (0.0138) | (0.0110) | (0.0171) | (0.0172) | (0.0172) |
| | | | | | | | | | |
| Covariates | Yes | Yes | Yes | Yes | Yes | Yes | Yes | Yes | Yes |
| Building FE | Yes | Yes | Yes | Yes | Yes | Yes | Yes | Yes | Yes |
| Census Block FE | Yes | Yes | Yes | Yes | Yes | Yes | Yes | Yes | Yes |
| Year–Month FE | Yes | Yes | Yes | Yes | Yes | Yes | Yes | Yes | Yes |
| Observations | 423,195 | 532,898 | 704,651 | 483,668 | 620,103 | 704,651 | 458,329 | 610,444 | 704,651 |
| Overall $R^2$ | 0.92228 | 0.91992 | 0.92069 | 0.90811 | 0.91185 | 0.92069 | 0.92107 | 0.92081 | 0.92069 |
| Within $R^2$ | 0.02964 | 0.04905 | 0.06955 | 0.05729 | 0.07517 | 0.06948 | 0.04925 | 0.07128 | 0.06953 |

Notes:

- Each column reports a separate pooled DID specification using the indicated comparison group and common reference group. Three-way interactions test whether the treatment effect differs between the comparison and reference groups. All models include fixed 2021 Just Value, building age at sale, distance to the coast, and building, Census Block, and year–month fixed effects.
- Standard errors are reported in parentheses and are two-way clustered by building and year–month.
- *** $p < 0.001$, ** $p < 0.01$, * $p < 0.05$, † $p < 0.1$.

**Table 3: Policy Anticipation and Monthly Four-Period DID Estimates**

| **DV: natural logged sales price** | | |
|---|---|---|
| | **Baseline** | **Adjusted** |
| Immediate Post-Surfside × Treated | −0.0270 | −0.0300* |
| | (0.0146) | (0.0148) |
| Policy Anticipation × Treated | −0.0371** | −0.0374** |
| | (0.0127) | (0.0129) |
| Post-SB4D × Treated | −0.0601*** | −0.0628*** |
| | (0.0075) | (0.0078) |
| 2021 Fixed Just Value (log) | | 0.2873*** |
| | | (0.0414) |
| Building Age at Sale | | 0.0045 |
| | | (0.0073) |
| Distance to Coast (log) | | 4.558 |
| | | (2.765) |
| | | |
| Building Fixed Effects | Yes | Yes |
| Census Block Fixed Effects | Yes | Yes |
| Year–Month Fixed Effects | Yes | Yes |
| Observations | 734,204 | 704,651 |
| Overall $R^2$ | 0.9147 | 0.9207 |
| Within $R^2$ | 0.0010 | 0.0695 |
| | | |
| Wald test (p-value):<br>$H_0$: Immediate Post-Surfside = Policy Anticipation | 0.637 | 0.737 |
| Wald test (p-value):<br>$H_0$: Policy Anticipation = Post-SB4D | 0.148 | 0.118 |
| Wald test (p-value):<br>$H_0$: Immediate Post-Surfside = Post-SB4D | 0.019 | 0.024 |

Notes:

- The omitted reference period is Pre-Surfside (January 2020–May 2021). Immediate Post-Surfside denotes June–September 2021. Policy Anticipation denotes October 2021–May 2022, beginning with the October 2021 release of the Florida Bar RPPTL Condominium Law and Policy Life Safety Advisory Task Force report. Post-SB4D denotes June 2022–December 2024.
- Standard errors are reported in parentheses and are two-way clustered at the building and year–month levels.
- *** $p < 0.001$; ** $p < 0.01$; * $p < 0.05$.

**Table 4: Monthly Three-Period DID Estimates, 12-Month Pre-Surfside Period**

| **DV: natural logged sales price** | | |
|---|---|---|
| | **Baseline** | **Adjusted** |
| Post-Surfside × Treated | −0.0298** | −0.0313*** |
| | (0.0088) | (0.0087) |
| Post-SB4D × Treated | −0.0575*** | −0.0608*** |
| | (0.0071) | (0.0075) |
| 2021 Fixed Just Value (log) | | 0.2899*** |
| | | (0.0405) |
| Building Age at Sale | | 0.0041 |
| | | (0.0067) |
| Distance to Coast (log) | | 3.966 |
| | | (2.944) |
| | | |
| Building Fixed Effects | Yes | Yes |
| Census Block Fixed Effects | Yes | Yes |
| Year–Month Fixed Effects | Yes | Yes |
| Observations | 666,574 | 638,746 |
| Overall $R^2$ | 0.9180 | 0.9244 |
| Within $R^2$ | 0.0009 | 0.0724 |
| | | |
| Linear pre-trend test (p-value) | 0.166 | 0.140 |
| Wald test (p-value): $H_0$: Post-Surfside = Post-SB4D | 0.016 | 0.012 |

Notes:

- The omitted (reference) period is Pre-Surfside (June 2020–May 2021). Post-Surfside denotes June 2021–May 2022, and Post-SB4D denotes June 2022–December 2024.
- Standard errors are reported in parentheses and are two-way clustered at the building and year–month levels.
- *** $p < 0.001$; ** $p < 0.01$; * $p < 0.05$.

**Figure 1: Monthly Event-Study Estimates**

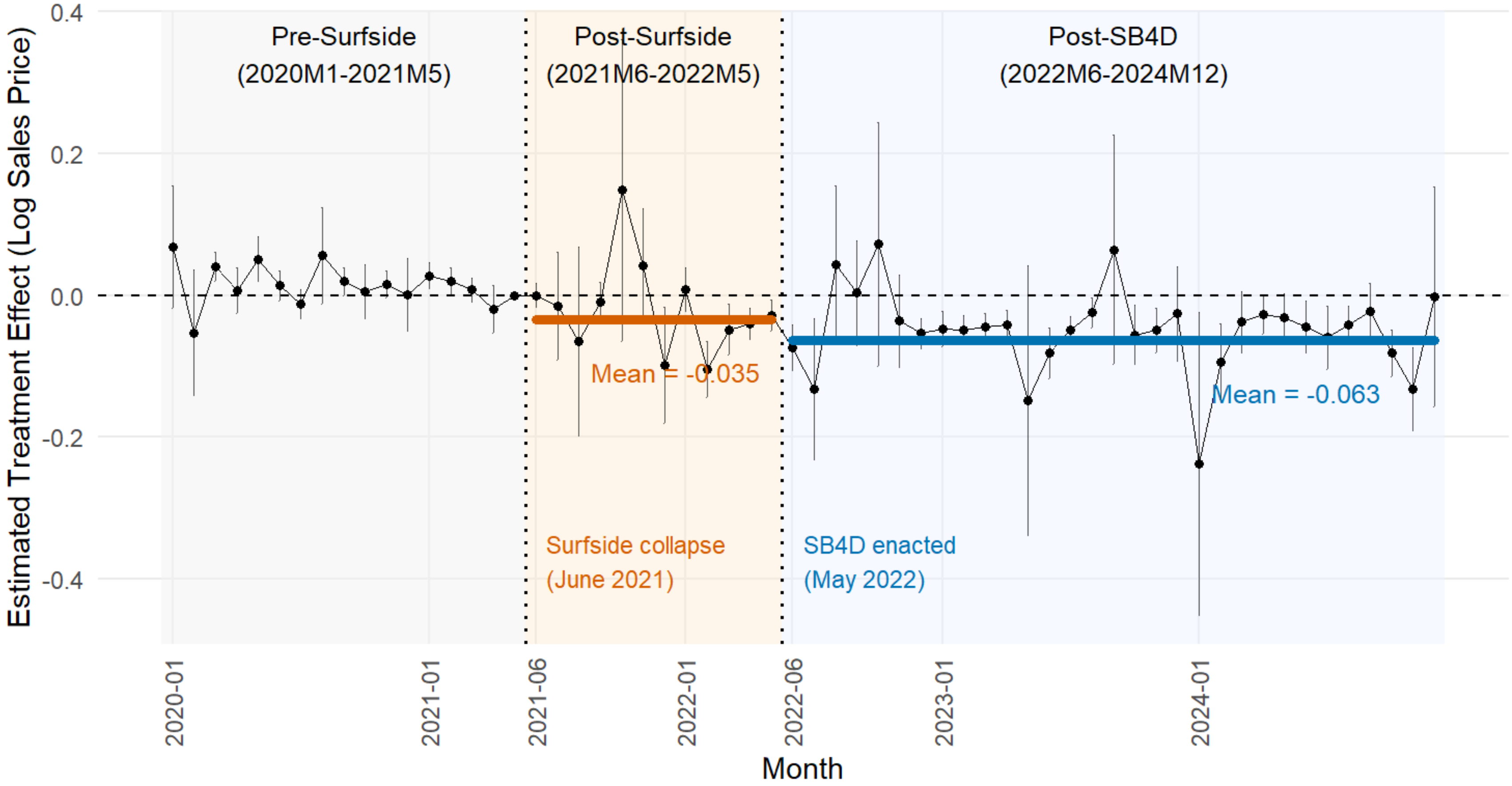

- Monthly treatment effects are estimated relative to May 2021, the month preceding the Surfside collapse.
- Vertical bars denote 95% confidence intervals based on standard errors two-way clustered by building and year-month.
- The adjusted specification includes fixed 2021 Just Value, building age at sale, and distance to coast.
- Pre-Surfside estimates provide a visual assessment of the parallel-trends assumption.
- Colored horizontal lines denote the average treatment effects estimated by the three-period DID specification.

**Figure 2: Monthly Event-Study Estimates, 12-Month Pre-Surfside Period**

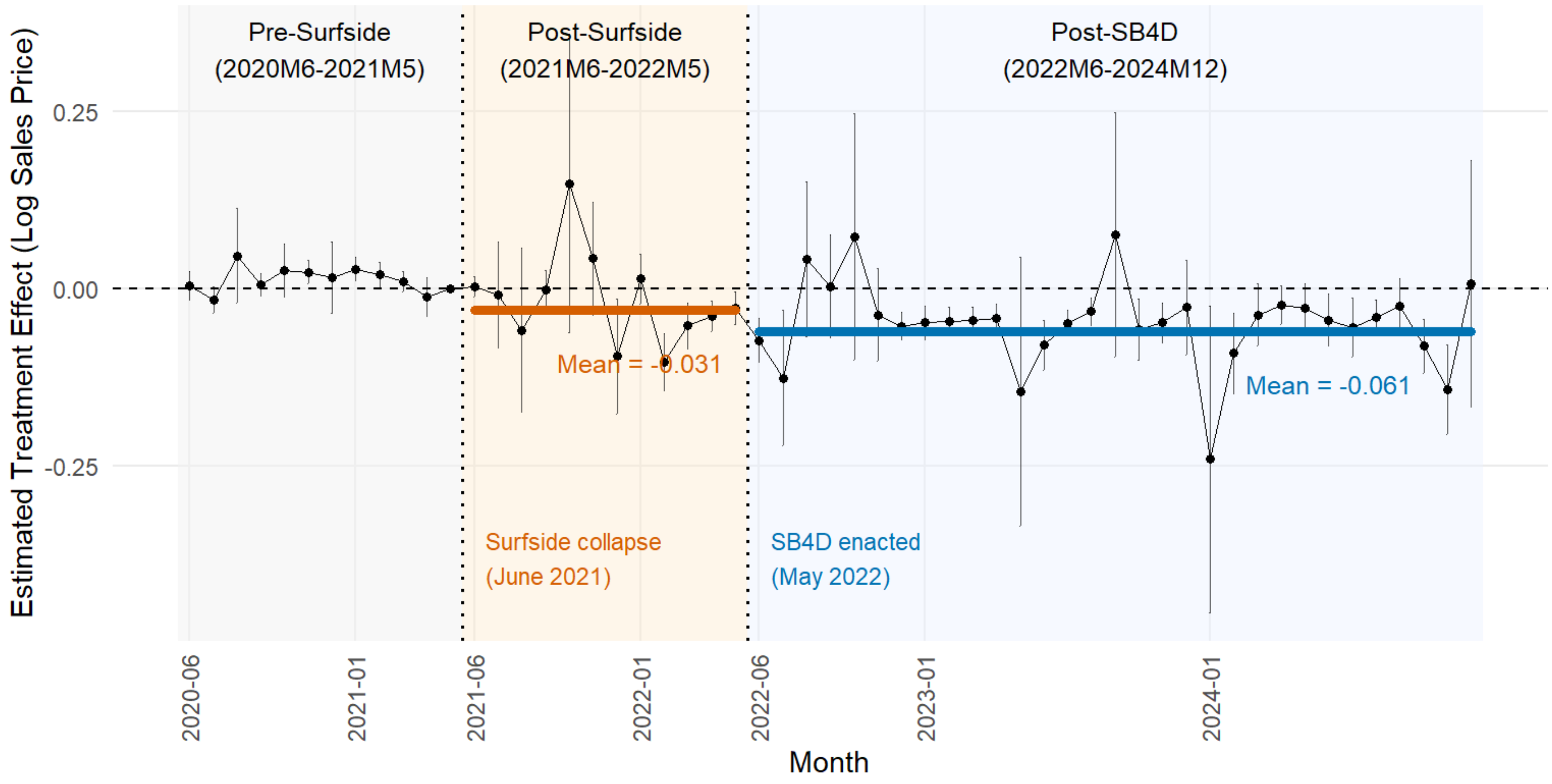

- Monthly treatment effects are estimated relative to May 2021, the month preceding the Surfside collapse.
- The sensitivity specification restricts the pre-Surfside period to June 2020-May 2021.
- Vertical bars denote 95% confidence intervals based on standard errors two-way clustered by building and year-month.
- The adjusted specification includes fixed 2021 Just Value, building age at sale, and distance to coast.
- Pre-Surfside estimates provide a visual assessment of the parallel-trends assumption.
- Colored horizontal lines denote the average treatment effects estimated by the corresponding three-period DID specification.